\documentclass[preprint,12pt]{aastex}

\def\snclus{J1004+411}
\def\abclus{ACO~1689}

\def\rs{r_{\hbox{\tiny{s}}}}
\def\rhol{\rho_{\hbox{\tiny{l}}}}
\def\rhog{\rho_{\hbox{\tiny{g}}}}

\def\zl{z_{\hbox{\tiny{L}}}} \def\zs{z_{\hbox{\tiny{S}}}}
\def\fls{c_{\hbox{\tiny{LS}}}}
\def\Dl{d_{\hbox{\tiny{OL}}}} \def\Ds{d_{\hbox{\tiny{OS}}}}
\def\Dls{d_{\hbox{\tiny{LS}}}}

\def\Omegam{\Omega_{\hbox{\tiny{m}}}}
\def\kpc{{\rm\;kpc}}
\def\munit{10^9M_\odot\;\rm kpc^{-2}}

\def\btheta{\hbox{\boldmath{$\theta$}}_{\hbox{\tiny{I}}}}
\def\bbeta{\hbox{\boldmath{$\theta$}}_{\hbox{\tiny{S}}}}

\begin{document}

\title{Two strong-lensing clusters \\
       confront universal dark-matter profiles}

\author{Prasenjit Saha and Justin I. Read}
\affil{Institute for Theoretical Physics, University of Z\"urich, \\
       Winterthurerstrasse 190, 8057 Z\"urich, Switzerland}
\author{Liliya L.R. Williams}
\affil{Department of Astronomy, University of Minnesota, \\
       116 Church Street SE, Minneapolis, MN 55455}

\begin{abstract}
In the galaxy clusters SSDS~\snclus\ and \abclus, a large number of
multiply-imaged background objects have recently been observed.  We
use these data to map the projected mass distribution in the inner
regions of these clusters.  The source redshifts in these clusters
span a large range, which eliminates the degeneracies that plague
nearly all lensing work.  As a result the mass maps are very
well-constrained, despite very weak prior assumptions.  \abclus\ 
lenses so many objects that we can afford to map it twice using
disjoint sets of images, thus verifying our internal error estimates.
We then deproject the mass maps (pretending for this purpose that they
are spherical) and obtain inner profiles consistent with $\rho\propto r^{-1}$
and indistinguishable from recent cold dark matter simulations.
Assuming that baryons make only a small difference to the profile outside
$\sim10\kpc$, these results support the prediction of a
universal inner profile for dark matter structures, independent of any
parametrization of that profile.
\end{abstract}


\section{Why inner profiles of clusters?}

A key prediction of heirarchical clustering dominated by cold dark
matter is that virialized structures have a universal density profile.
Although there is still no first-principles derivation of a universal
profile, $N$-body simulations and phenomenological models both
consistently indicate that such a profile exists.

The best-known candidate for a universal profile remains
the original proposal
\begin{equation}
\rho \propto (r/\rs)^{-1} (1+r/\rs)^{-2}
\end{equation}
from \cite{1997ApJ...490..493N}, but it is gradually being superseded.
In recent simulations ---see for example, Fig.~3 of
\cite{2004MNRAS.349.1039N}--- NFW remains a reasonable fit to
individual simulated halos, but an ensemble of halos shows the
steepness changing gradually with no distinct $\rs$.  \cite{graham}
discuss alternative parametrizations.  The smallest resolved scales
in current dark-matter simulations have $\rho\sim r^{-1.2\pm0.2}$
\citep{2004MNRAS.353..624D}.
The total mass profile would be somewhat steeper at the smallest
radii, because of the infall of baryons and the response of the
dark-matter to the baryons.  The latter effect is generally modelled
as adiabatic contraction
\citep{1986ApJ...301...27B,2004ApJ...616...16G}.

These results motivate research on the inner profiles of real
clusters.

\section{Why \snclus\ and \abclus?}

Testing predicted universal profiles against observations seems
straightforward at first.  Well-known strategies are to
\begin{enumerate}
\item put observed velocity dispersions through the Jeans equation to
infer a density profile, or
\item fit a mass model to lensing data, or
\item invert an observed x-ray profile through the hydrostatic
equilibrium equation.
\end{enumerate}
The first two approaches, and to a lesser extent the third, have all
been followed in the literature, but the conclusions remain unclear.
For example: \cite{2006MNRAS.367.1463L} using method (1) find NFW
profiles satisfactory, whereas \cite{2004ApJ...604...88S} using a
combination of (1) and (2) find inner profiles significantly shallower
than NFW.

Evidently there are major uncertainties involved in extracting a mass
profile. It is useful to get some insight into what the uncertainties
are, and hence how robust different methods are likely to be.  By
considering the basic equations involved, we can identify problematic
operations that either amplify noise or introduce systematic errors.
Let us call them gremlins.

The first gremlin appears in deprojecting an observed quantity, via
the well-known Abel integral
\begin{equation}
\rho(r) = -\frac1\pi\int_r^\infty
          \frac{d\Sigma(R)}{dR} \frac{dR}{\sqrt{R^2-r^2}} .
\label{abel}
\end{equation}
In this case, differentiating an observed $\Sigma(R)$ amplifies the
noise.  Smoothing $\Sigma(R)$ will suppress the noise of course, but
will replace it with a systematic error.  However, the integral tends
to suppress the noise somewhat, so this gremlin is comparatively
benign.

Next, consider Jeans equation for spherical systems
\begin{equation}
GM(r) = \frac{r^2}\rhol \frac d{dr} (\rhol\sigma_r^2)
      - 2r(\sigma_r^2-\sigma_t^2)
\label{jeans}
\end{equation}
relating the luminous density $\rhol$ and velocity dispersions to
the mass profile.  Since $\sigma_r$ and $\sigma_t$ cannot be
directly observed, they need to be related to the projected dispersion
somehow.  This introduces a second gremlin: assumptions about
anisotropy produce a systematic error; deriving the anisotropy from
higher velocity moments \citep{2006MNRAS.367.1463L} introduces noise.
The derivative in Eq.~(\ref{jeans}) is a third gremlin.

For x-ray emitting gas in hydrostatic equilibrium, the temperature
profile $T(r)$ and gas density profile $\rhog(r)$ are related to the
total mass profile by
\begin{equation}
GM(r) = \frac{kT}{\mu m_p} r^2
        \left( \frac1T \frac{dT}{dr} + \frac1{\rhog} \frac{d\rhog}{dr} \right)
\label{hydros}
\end{equation}
where $\mu_p$ is the proton mass and $\mu$ the particle weight.  Here
the first gremlin appears in deprojecting for $T(r)$ and $\rhog(r)$,
while the third gremlin appears in another guise in the derivatives
of $T$ and $\rhog$.

Lensing is free of the second and third gremlins.  Moreover it
provides one simple and robust constraint on mass profiles.  Central
images are extinguished by an isothermal or steeper cusp
\citep{1998MNRAS.296..800E}, and the fact that clusters commonly host
central images implies that clusters are shallower than isothermal
\citep{2000ApJ...538..528S}.

Lensing, however, introduces a new gremlin, namely lensing
degeneracies.  Consider the
equation for the arrival time\footnote{Eq.~(\ref{arriv}) is Eq.~(2.6)
from \cite{1986ApJ...310..568B}, except that we have introduced three
new symbols: $\nabla^{-2}$ is an operator that solves Poisson's
equation in 2D, $\fls=\Dls/\Ds$, while $\kappa=4\pi\Dl\Sigma$ is the
surface density in units of the critical density for sources at
infinity.  Note that Blandford \& Narayan set $G=c=1$.}
\begin{equation}
\tau(\btheta;\bbeta) = {\textstyle\frac12} (\btheta-\bbeta)^2
                - 2 \fls \nabla^{-2} \kappa(\btheta)
\label{arriv}
\end{equation}
which represents the time delay for a virtual photon\footnote{Real
photons take paths for which $\tau$ is a maximum, minimum, or saddle
point (Fermat's principle).} originating from a source at sky position
$\bbeta$ and getting deflected at the lens so that the observer sees
it coming from sky-position $\btheta$.  The first term on the right of
Eq.~(\ref{arriv}) is the geometrical path difference between a
deflected and undeflected photon trajectory, and the last term is the
Shapiro time delay.  The whole equation is dimensionless: $\tau$ is a
scaled time delay and $\kappa$ is a scaled surface density, the
coordinates are in radians, while $\fls\sim1$ is a ratio of distances
depending on lens and source redshifts.  The gradient of $\tau$ with
respect to $\btheta$ gives the lens equation, and the matrix of second
derivatives is the inverse magnification.  Thus, image positions in
strong lensing constrain the first derivative of $\tau$ while shear or
magnification in weak or strong lensing constrain the second
derivative.  These derivatives are not gremlins, because no
differentiation of data is involved.  But let us rewrite the equation
as
\begin{equation}
\tau = 2\,\nabla^{-2} (1 - \fls\kappa) - \btheta\cdot\bbeta
\label{arriv2}
\end{equation}
where we have expanded the geometrical part and then discarded a
$\bbeta^2$ term since it has no optical effect.  If $\fls$ is a
constant (that is, there is only one source redshift) then we may
rescale Eq.~(\ref{arriv2}) by some positive factor $\lambda$:
\begin{equation}
\tau \leftarrow \lambda \tau, \quad
\bbeta \leftarrow \lambda \bbeta, \quad
(1-\fls\kappa) \leftarrow \lambda (1-\fls\kappa).
\label{rescale}
\end{equation}
Eq.~(\ref{rescale}) amounts to redefining the contour spacing in
Blandford \& Narayan's Fig.~2, without changing the figure --- the
observables don't change, but the $\kappa$ profile becomes steeper (if
$\lambda>1$ or shallower (if $\lambda<1$).  This is our fourth, and
perhaps most vicious, gremlin. The case where $\lambda$ is constant is
well known as a steepness degeneracy \citep{1985ApJ...289L...1F}, but
in fact $\lambda$ need not even be constant
\citep{2000AJ....120.1654S,2003ApJ...582....2Z,kann}.  Having a range
of source redshifts breaks the degeneracies
\citep{1998AJ....116.1541A}; having number counts of weakly lensed
objects breaks the main steepness degeneracy
\citep{2002A&A...386...12D}.

The above arguments indicate that lens reconstructions with special
attention to breaking lensing degeneracies are likely to provide the
most robust mass profiles.  The clusters which have been so analyzed
are ACO~2218 \citep{1996ApJ...471..643K,1998AJ....116.1541A} ACO~370
\citep{1998MNRAS.294..734A}, \abclus\ 
\citep{2002A&A...386...12D,2005ApJ...621...53B,2005MNRAS.362.1247D,
2006ApJ...640..639Z,hsp06}, and \snclus\ \citep{2005ApJ...629L..73S}.
The first two have double-peaked mass distributions, evidently from
ongoing mergers.  That leaves the last
two lensing clusters as worthy of special attention.

\section{Mass profiles}

We will now use the excellent recent strong-lensing observations by
\cite{2005ApJ...629L..73S} on \snclus\ and by
\cite{2005ApJ...621...53B} on \abclus\ to map the mass distribution in
the inner regions of these clusters.  As we noted above, these data
have already made into models by the observing authors and others.
Our work, however, goes further than previous models because
(i)~rather than fitting one or a few models to the data, we generate
large ensembles of models exploring the possible mass distributions
that can reproduce the data; (ii)~we compare results from independent
data sets to test for consistency; and (iii)~we deproject the surface
density profiles non-parametrically for a better comparison with
simulations.

Our technique is implemented in the {\em PixeLens\/} code, originally
developed for galaxy lenses \citep{2004AJ....127.2604S}, but now
enhanced to use multiple source-redshifts.  Given the multiple-image
data, the program generates an ensemble of 400 models that each
(a)~reproduce the lensing data exactly, and (b)~are consistent with a
given prior on the mass map.  The model ensemble automatically
provides Bayesian uncertainties on the mass map or any derived
quantity.
We assume a concordance cosmology ($K=0,\Omegam=0.3,w=-1$) with
$H_0^{-1}=14\rm\,Gyr$.  Cosmological parameters enter into the the
mass and distance scales, but not the profiles themselves.

The prior we assume is the same as in \cite{2004AJ....127.2604S}
---basically it requires that that the surface density be non-negative
and centrally concentrated, with substructure allowed--- but with one
difference: we remove the requirement that $\Sigma$ be steeper than
$R^{-0.5}$ (which was motivated by galaxy dynamics) and require only
that the cylindrically-averaged $\Sigma$ decrease with $R$.  We
measure $R$ from the center of the brightest cluster galaxy,
so the mass peak is modeled to coincide with the light peak.
In some lensing clusters mass and light peaks have been inferred to be
offset tens of kpc \citep{1998MNRAS.294..734A,1998AJ....116.1541A},
but in \snclus\ and \abclus\ the central lensed images indicate that
any such offset is small.

The requirement of exact fits to the image data leaves {\em
PixeLens\/} currently unable to handle all the multiple-image data on
\abclus.  (The problem is accumulation of roundoff error.)
Nevertheless, we can make a virtue of this by splitting the data into
two independent sets.  Accordingly, we present three mass reconstructions:
\begin{enumerate}
\item \snclus\ ($\zl=0.68$) reconstructed using 13 images coming from
4 sources ($\zs=1.73$ to 3.33),
\item \abclus\ ($\zl=.18$) reconstructed using 30 images of 9 sources
($\zs=1.74$ to 4.52), and
\item \abclus\ again, reconstructed using 28 images of 7 other
sources ($\zs=1.57$ to 5.16).
\end{enumerate}

\begin{figure}
\epsscale{0.3}  \plotone{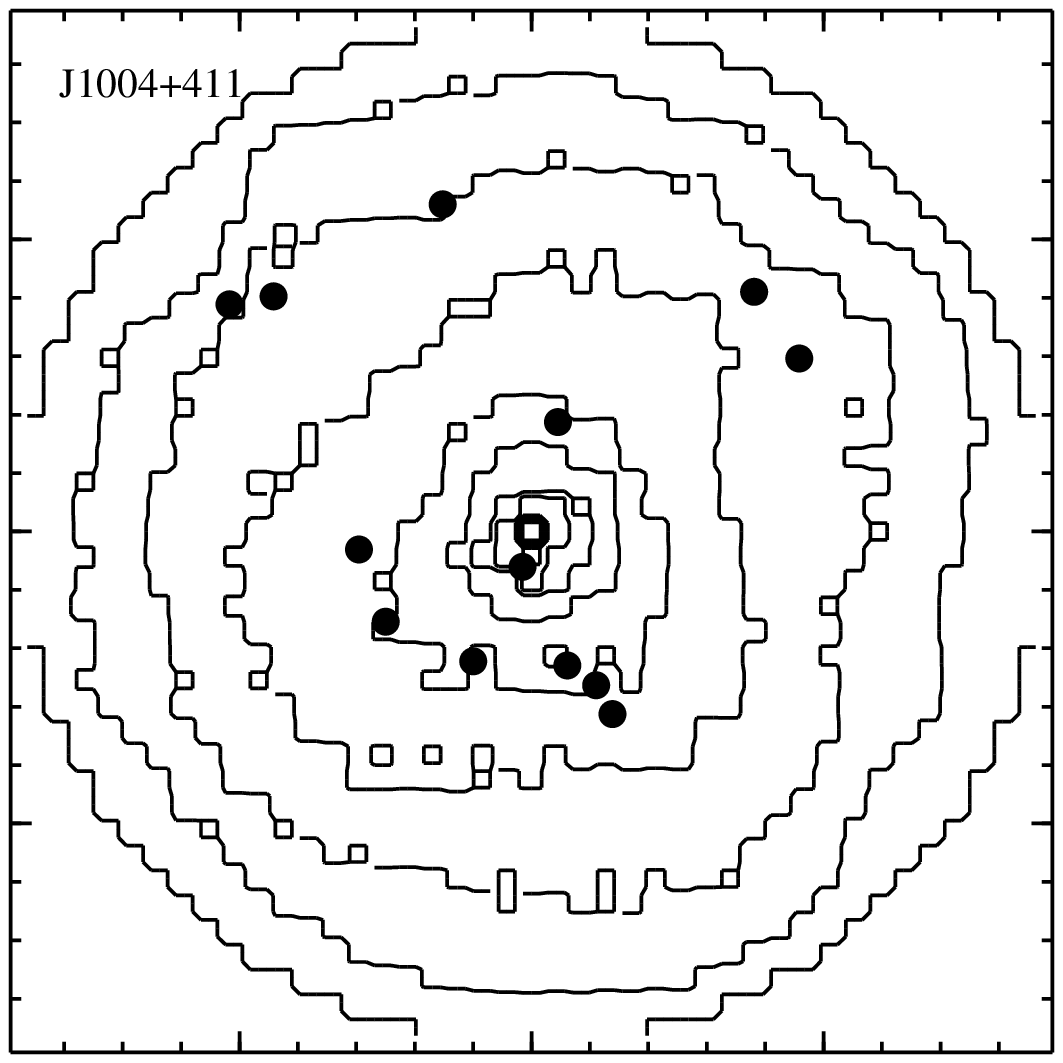}
\epsscale{0.31} \plotone{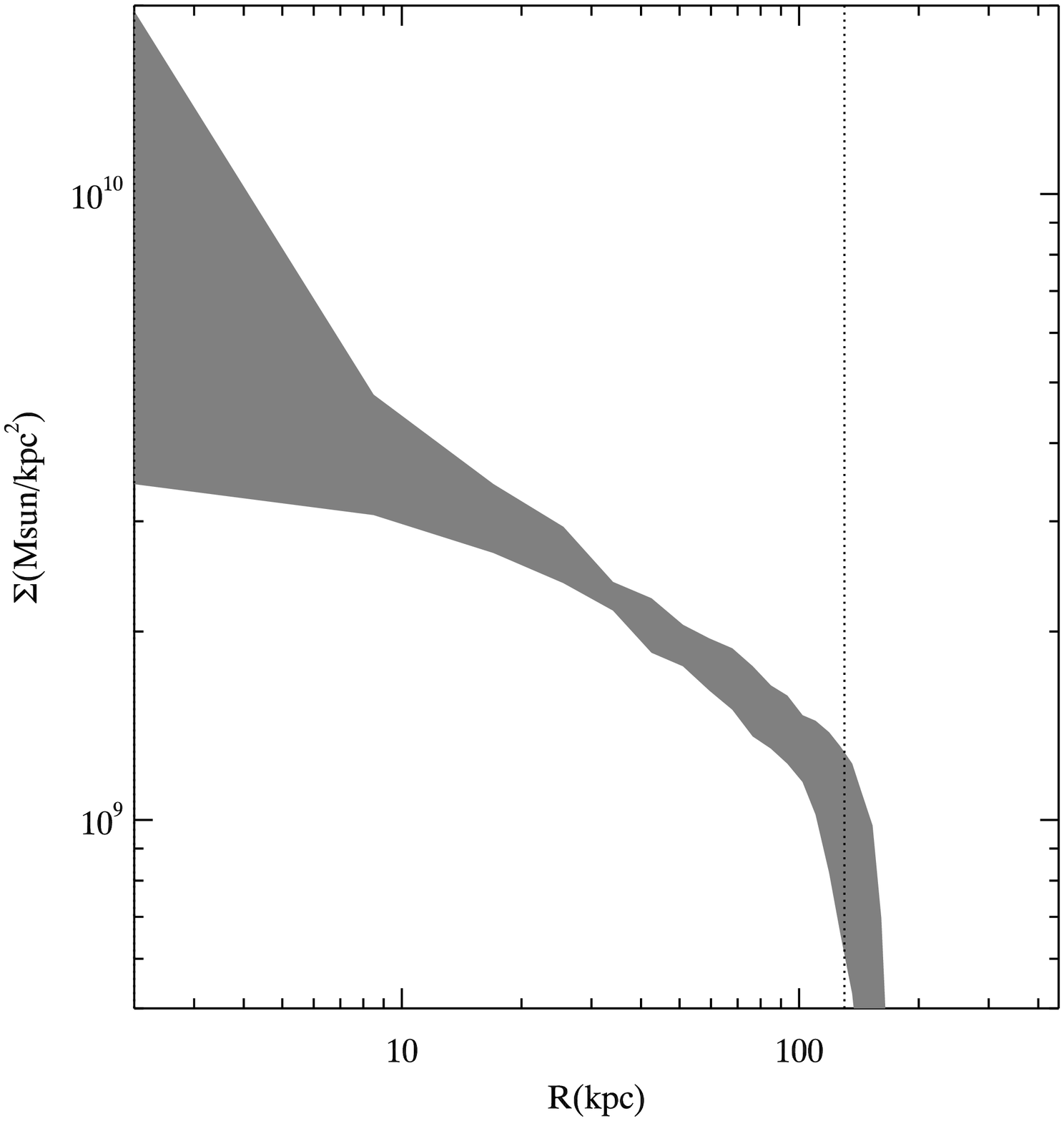} \plotone{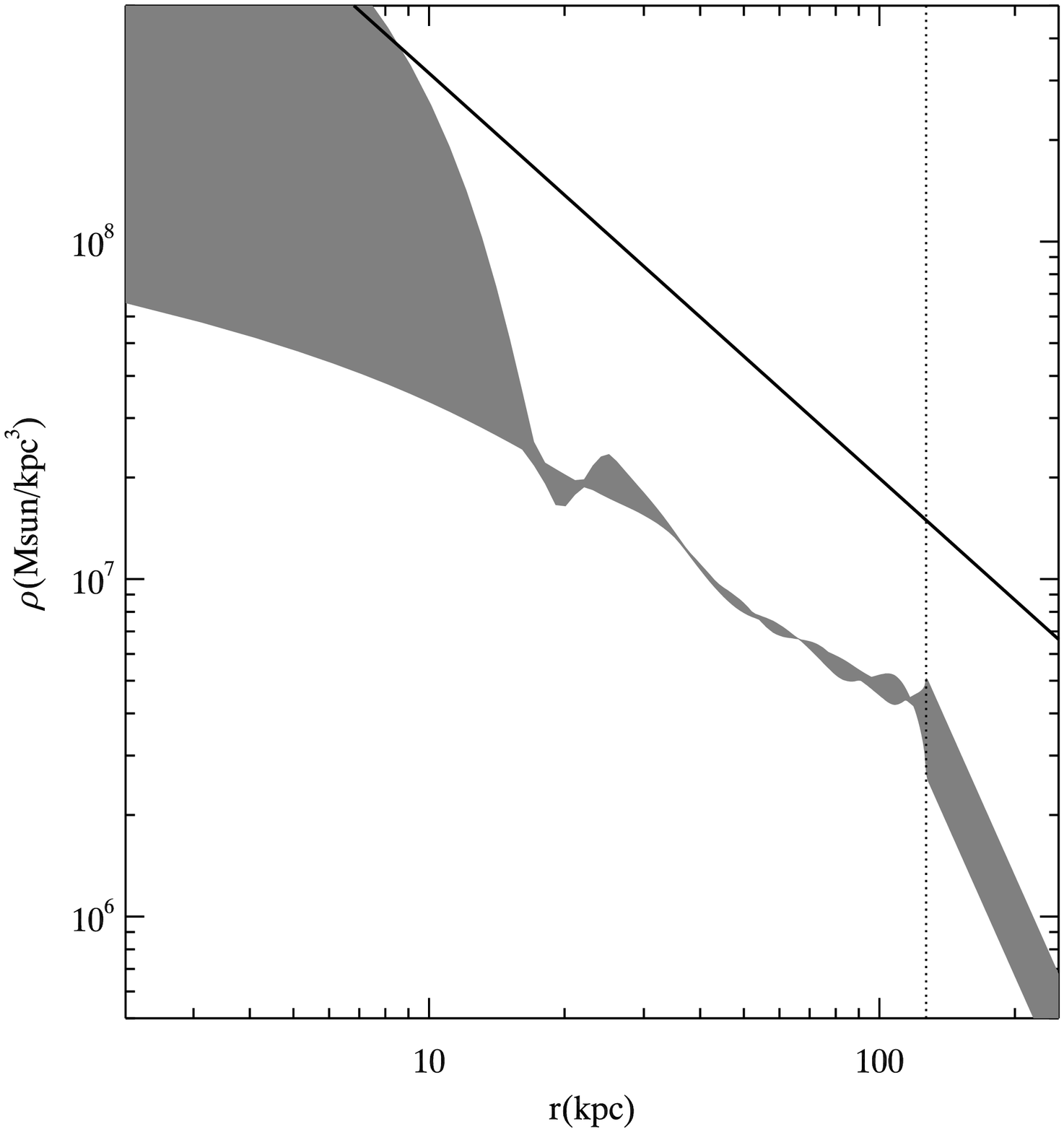}
\goodbreak
\epsscale{0.3}  \plotone{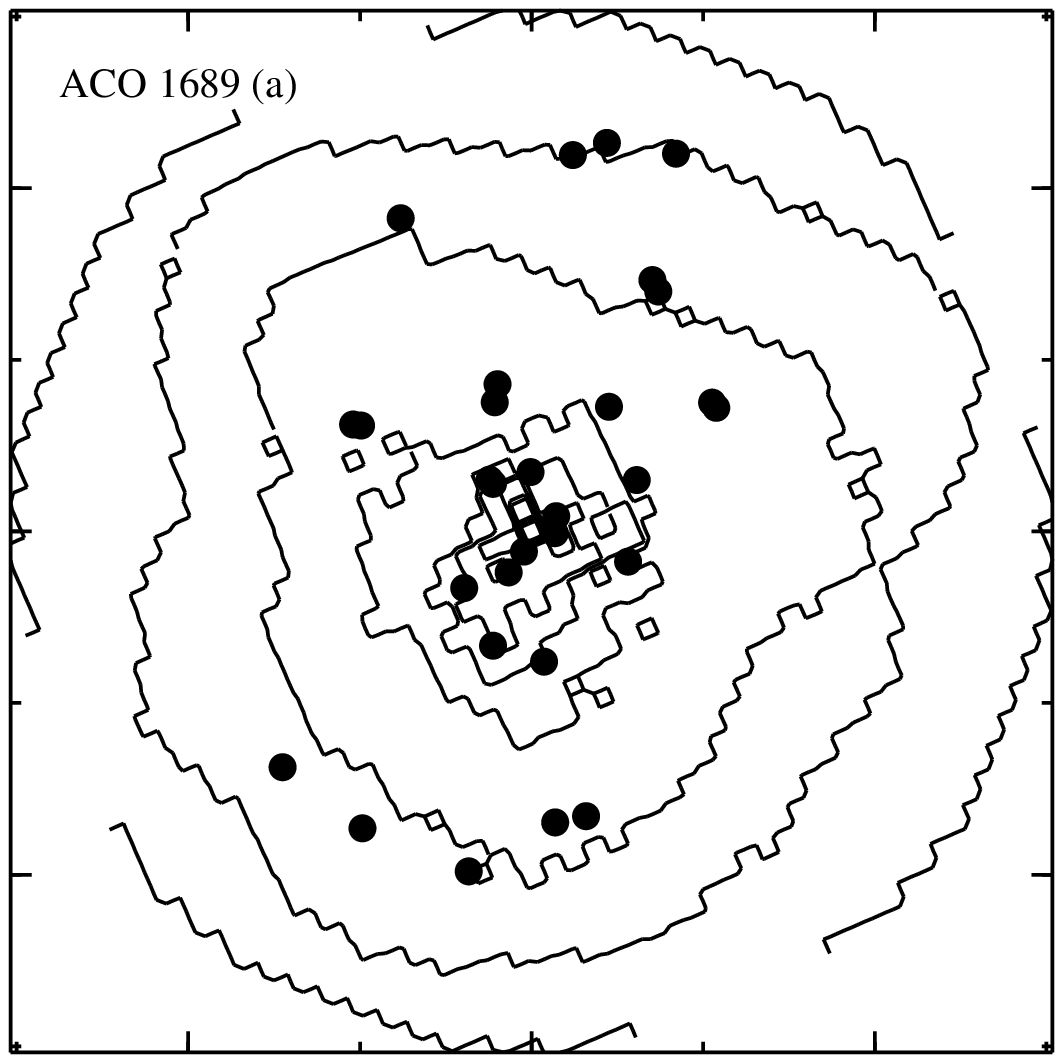}
\epsscale{0.31} \plotone{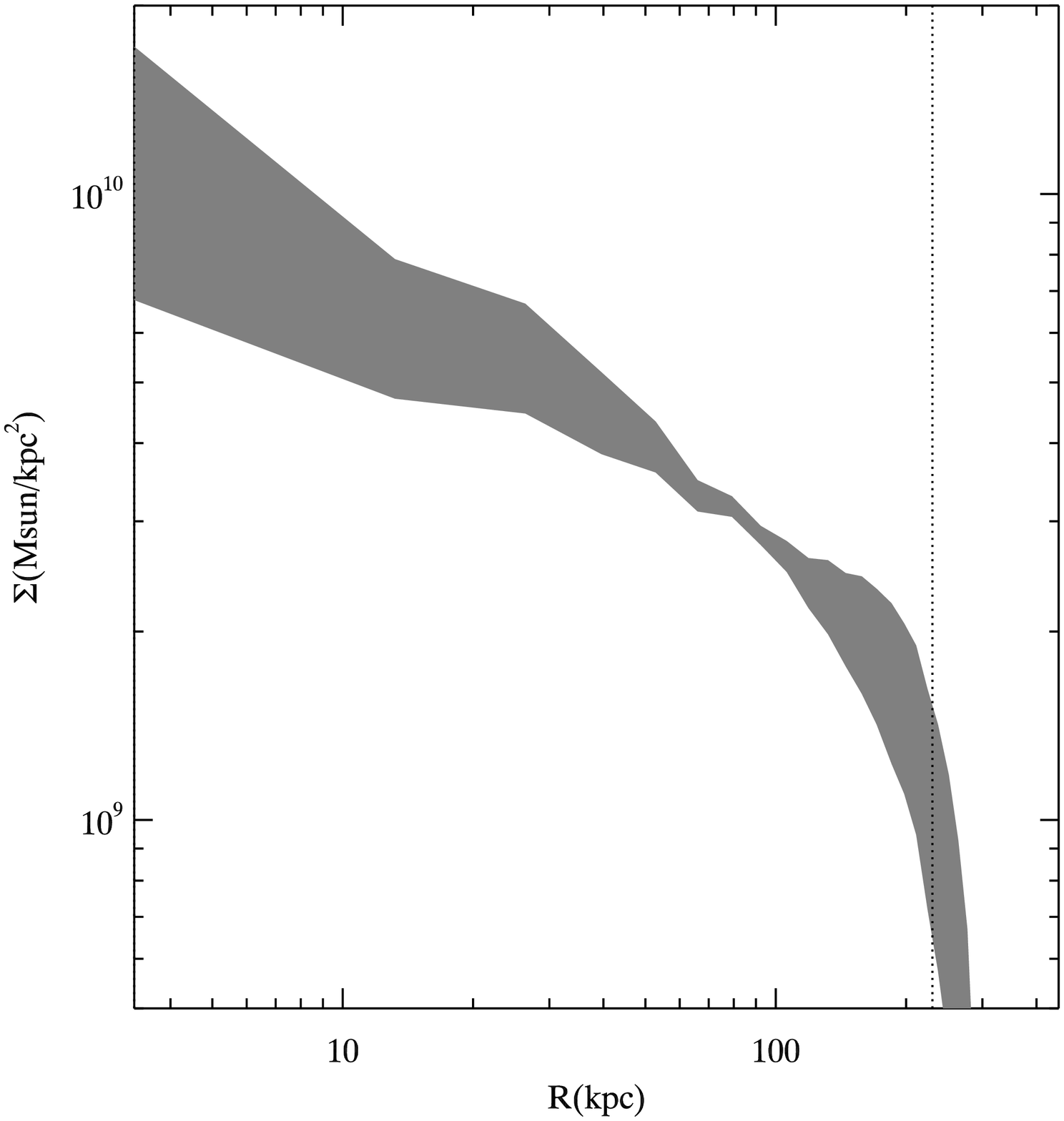} \plotone{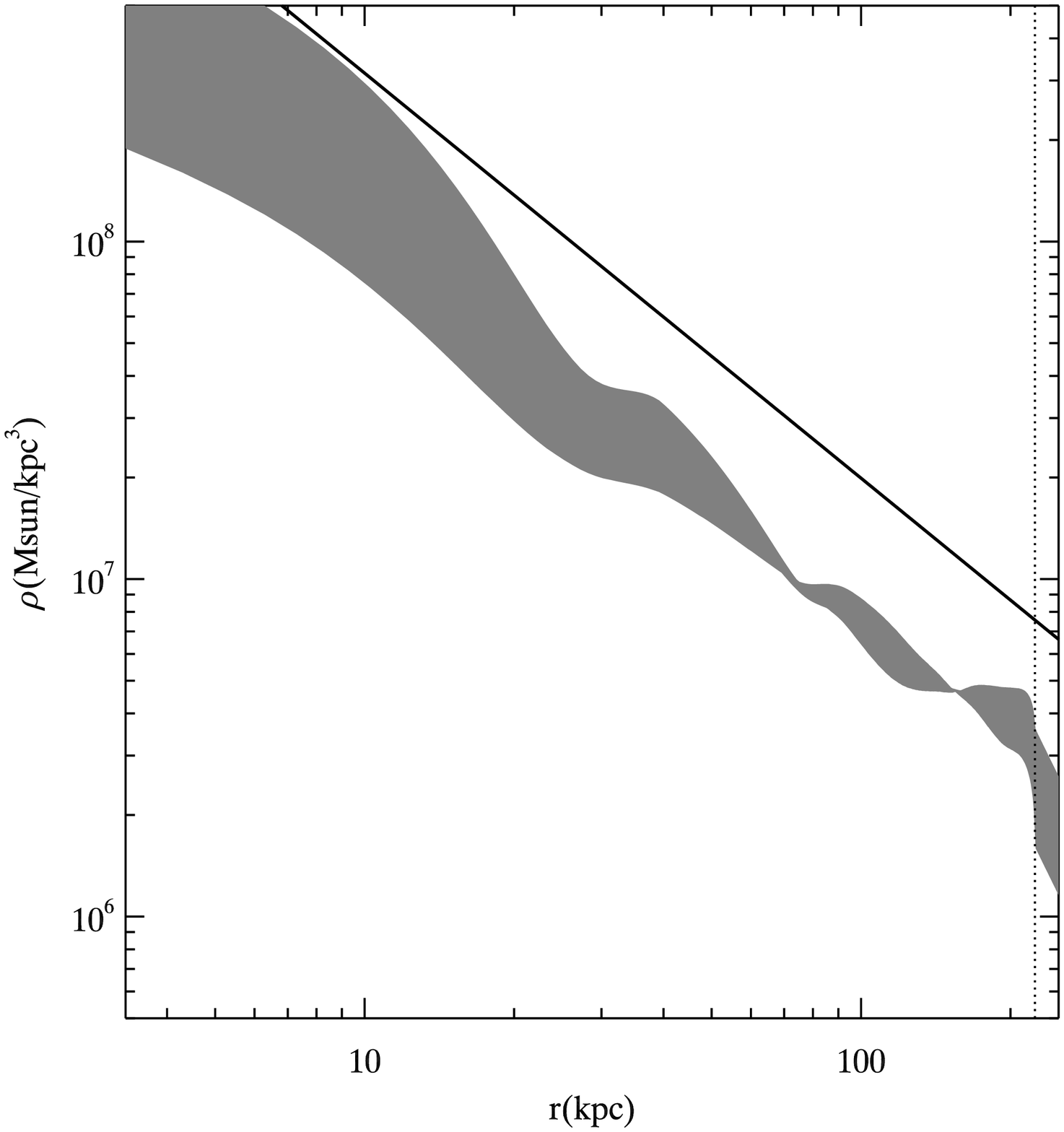}
\goodbreak
\epsscale{0.3}  \plotone{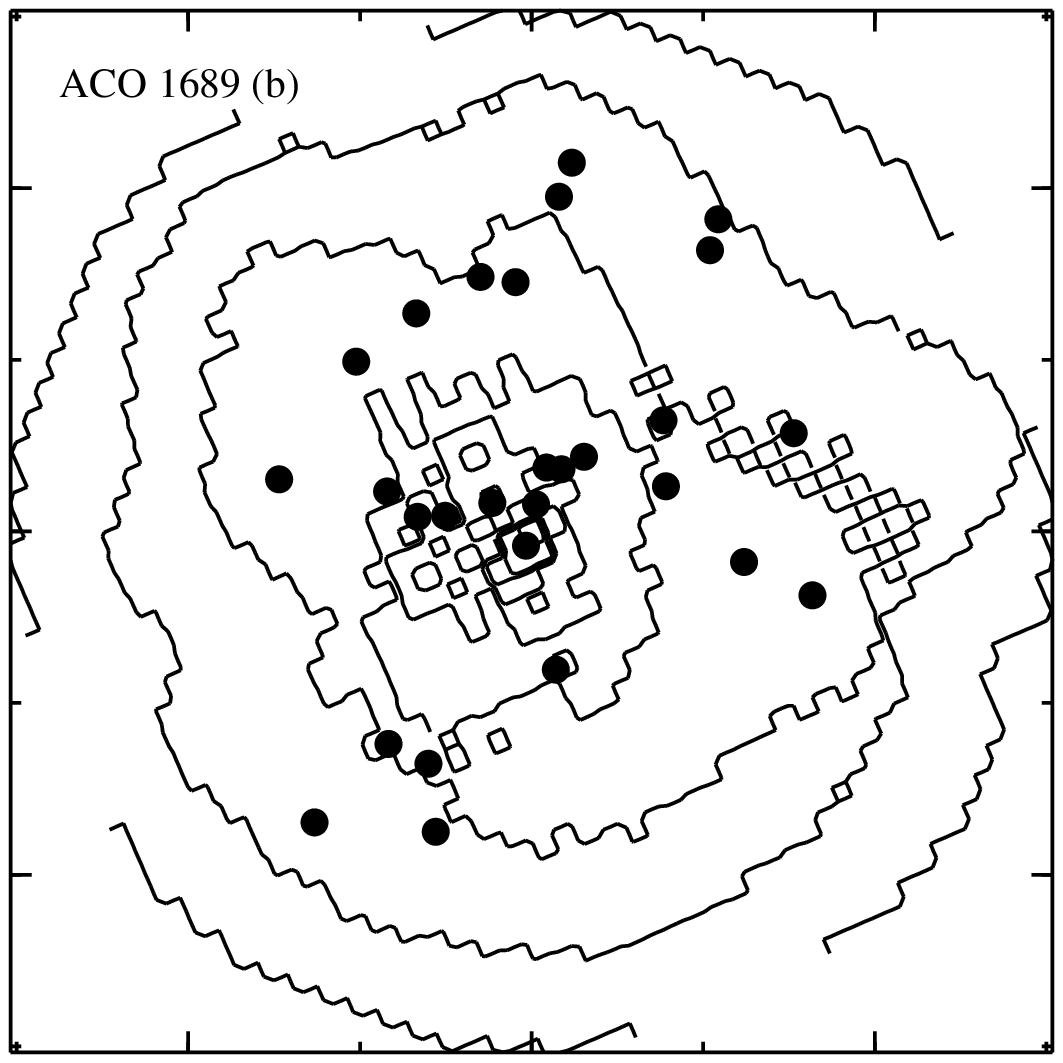}
\epsscale{0.31} \plotone{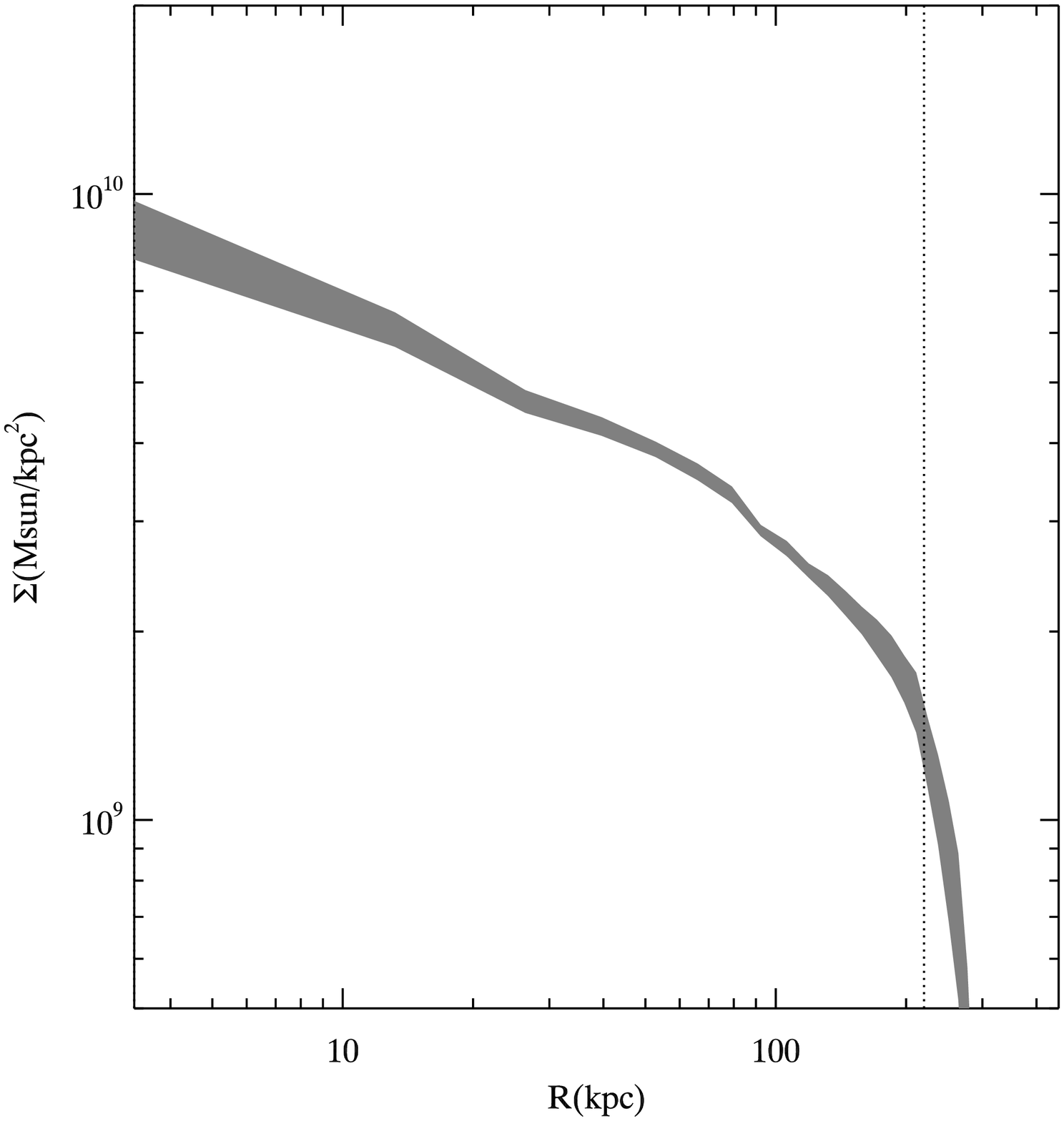} \plotone{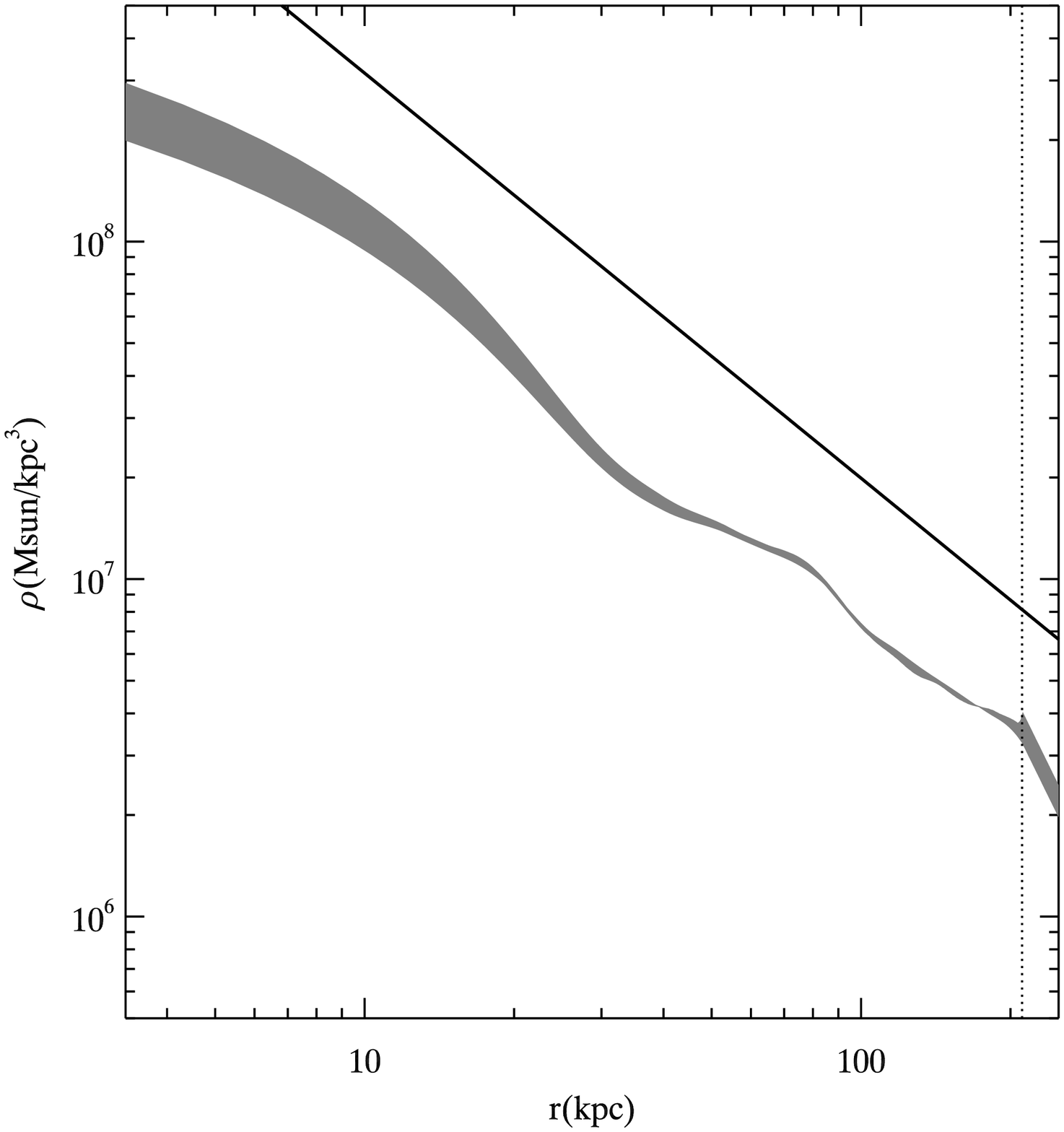}
\caption{Mass maps and profiles for \snclus\ and for \abclus. {\em Left
column:\/}~Ensemble-average mass maps together with the lensed image
positions. Panels are oriented as North-up/East-left. Ticks correspond
to $20\kpc$ for \snclus\ and $100\kpc$ for \abclus.  Contour steps are
$0.5\times\munit$ for \snclus\ and $\munit$ for \abclus.  {\em Middle
column:\/}~$\Sigma(R)$ for the mass maps in in the left column,
together with 90\% uncertainties. The vertical lines to the right show
$R$ for the outermost image.  {\em Right column:\/}~Density profiles,
derived by deprojecting the profiles from the middle column. The
sloped line shows $\rho\propto r^{-1.2}$.}
\label{all}
\end{figure}

Figure \ref{all} packs all the results of this paper.  To the left
we show three ensemble-average mass maps.
Both clusters appear fairly round, but have some substructure,
such as a feature to the NE in \abclus.  In this paper we
concentrate on radial dependence, leaving substructure for later work.

The middle column of the figure shows the cylindrically-averaged
surface density $\Sigma(R)$, with 90\%-confidence uncertainties
derived from the
model-ensemble.  In each case we see a shallow and tightly constrained
profile in the inner region, and moreover the two profiles for
\abclus\ from independent image-sets agree. (The profile beyond the
outermost image is to be disregarded; there {\em PixeLens\/} has no
data to work with and simply makes the profile fall smoothly to zero
at the edge of the map.)  The enclosed mass agrees with previous
strong-lensing estimates but is about a factor of two higher than
current models for the x-ray gas give \citep{2004ApJ...607..190A}.
The improvement upon our earlier reconstructions of \snclus\
\cite{2004AJ....128.2631W} using a single 4-image background object is
remarkable: then we concluded that the lens was consistent with NFW,
but even a profile as steep as $R^{-1.3}$ was not excluded; now simply
inspecting the upper middle panel of Fig.~\ref{all} shows that this
cluster is shallower than $1/R$.

To the right of the figure we show $\rho(r)$, derived from $\Sigma(R)$
using Eq.~(\ref{abel}).  To evaluate the numerical derivative and then
the integral, we fit a spline to $\Sigma$ up to the $R$ of the
outermost image and assumed $\Sigma\propto R^{-2}$ outside.  (The
result is not sensitive to the assumed outside slope, or to the
boundary between spline and $R^{-2}$, provided the latter is not
beyond the image region.)  Here again we show
ensemble-derived uncertainties, which in this case are probably too
small, because substructure will introduce systematic errors through
the derivative in Eq.~(\ref{abel}).  With this caveat, our constraints
on the steepness between 10\thinspace kpc and the outermost image are
as follows: \snclus\ is best fit by $\rho\propto r^{-0.9}$ but the
uncertainty bands shown allow $r^{-0.8}$ to $r^{-1.2}$, while \abclus\ is
best fit by $\rho\propto r^{-1.15}$ but the uncertainty bands allow
$r^{-1.0}$ to $r^{-1.25}$.  These values are typical of the inner
slopes in Table~3 of \cite{2004MNRAS.353..624D}.  The right column of
Fig.~\ref{all} is also comparable to their Fig.~4.
For \abclus\ we have also
attempted to model the change in the profile due to adiabatic
contraction, using estimated gas density from Fig.~9 of
\cite{2004ApJ...607..190A}, but the difference is smaller than our
estimated uncertainties.  It appears that baryons make a difference of
at most 10\%, which is of the same order as the substructure.
We conclude that, within current uncertainties, the inner slopes in
these lensing clusters and in simulations are indistinguishable.

\section{Discussion}

In this paper we have used recent strong lensing observations having
multiple source-redshifts
\citep{2005ApJ...629L..73S,2005ApJ...621...53B} to reconstruct the
inner density profiles of two clusters in a non-parametric way.  These
inner density profiles show remarkable agreement with simulations of
CDM halos.  Although we have disregarded the secondary questions of
baryons (not included in simulations at this scale) and substructure
(present in both the real Universe and simulated ones), our results
arguably confirm an important prediction of heirarchical structure
formation.

But why are the profiles the way they are?  There is still nothing
approaching a first-principles derivation of a universal profile,
though there are several interesting ideas.  The only exactly solved
case is self-similar spherical collapse, but that implies $\rho\propto
r^{-2.25}$ \citep{1984ApJ...281....1F,1985ApJS...58...39B} which is
too steep.  \cite{2000ApJ...538..528S} propose a cunning variant
involving merging self-similar proto-halos into a Matreshka-like
heirarchy, resulting in a composite halo with a shallow inner slope;
but their mechanism relies on undigested merging.  Several other
scenarios involving spherical collapse have also been studied, and
recently \cite{2006MNRAS.368.1931L} proposed a phenomenological 1D
model that produces an $r^{-1}$ inner slope and an $r^{-3}$ outer
slope.  Finally, one numerical result is especially intriguing: in CDM
collapse simulations the surrogate phase-space density
$\rho/\sigma^3\propto r^{-\alpha}$
\citep{2001ApJ...563..483T,2005ApJ...634..775B}. Combining this
relation with the Jeans equation gives $\rho\propto r^{-0.8}$
asymptotically for small $r$
\citep{2005ApJ...634..756A,2006EAS....20...33H,2005MNRAS.363.1057D}.
The total mass profile at very small $r$ would be steeper than this
because of the baryons, and anyway our results do not have the
resolution to test for this yet.  But it remains an interesting
prediction.


\bibliographystyle{apj}
\bibliography{ms.bbl}


\end{document}